\documentclass[twocolumn,pra,preprintnumbers,amsmath,amssymb]{revtex4-1}
\usepackage{ae,aecompl}

\usepackage{graphicx}
\usepackage[normalem]{ulem}
\usepackage{soul,xcolor}
\setstcolor{red}

\usepackage[latin1]{inputenc}
\usepackage[T1]{fontenc}

\usepackage{color}

\newcommand{\etal}{{\it et al.}}

\begin{document}

\def\pss#1#2#3{Phys.~Stat.~Sol.~{\bf #1},\ #2\ (#3)}
\def\apl#1#2#3{Appl.~Phys.~Lett.~{\bf #1},\ #2\ (#3)}
\def\jpb#1#2#3{J.~Phys.~B:~{\bf #1},\ #2\ (#3)}
\def\jpc#1#2#3{J.~Phys.~Chem.~{\bf #1},\ #2\ (#3)}
\def\jpcb#1#2#3{J.~Phys.~Chem. B~{\bf #1},\ #2\ (#3)}
\def\jcp#1#2#3{J.~Chem.~Phys.~{\bf #1},\ #2\ (#3)}
\def\cpl#1#2#3{Chem.~Phys.~Lett.~{\bf #1},\ #2\ (#3)}
\def\pr#1#2#3{Phys.~Rev~{\bf #1},\ #2\ (#3)}
\def\pra#1#2#3{Phys.~Rev.~A~{\bf #1},\ #2\ (#3)}
\def\prb#1#2#3{Phys.~Rev.~B~{\bf #1},\ #2\ (#3)}
\def\prc#1#2#3{Phys.~Rev.~C~{\bf #1},\ #2\ (#3)}
\def\prd#1#2#3{Phys.~Rev.~D~{\bf #1},\ #2\ (#3)}
\def\pre#1#2#3{Phys.~Rev.~E~{\bf #1},\ #2\ (#3)}
\def\rmp#1#2#3{Rev.~Mod.~Phys.~{\bf #1},\ #2\ (#3)}
\def\prl#1#2#3{Phys.~Rev.~Lett.~{\bf #1},\ #2\ (#3)}
\def\sci#1#2#3{Science~{\bf #1},\ #2\ (#3)}
\def\nat#1#2#3{Nature~{\bf #1},\ #2\ (#3)}
\def\apj#1#2#3{Astrophys.~J.~{\bf #1},\ #2\ (#3)}

\def\bea{\begin{eqnarray}}
\def\eea{\end{eqnarray}}
\def\be{\begin{equation}}
\def\ee{\end{equation}}
\def\etal{{\it et al.}}

\newcommand{\JCP}[1]{J. Chem. Phys. {\bf #1}}
\newcommand{\ApJ}[1]{Astrophys. J. {\bf#1}}
\newcommand{\CPL}[1]{Chem. Phys. Lett. {\bf #1}}
\newcommand{\PRA}[1]{Phys. Rev. A {\bf #1}}
\newcommand{\JPC}[1]{J. Phys. Chem. {\bf #1}}
\newcommand{\JPCA}[1]{J. Phys. Chem. A {\bf #1}}
\newcommand{\PRL}[1]{Phys. Rev. Lett. {\bf #1}}
\def\BE{\begin{equation}} \def\EE{\end{equation}}

\title{Measurements of trapped-ion heating rates with exchangeable surfaces in close proximity}
%\date{\today}
\author{D. A. Hite}
\author{K. S. McKay}
\author{S. Kotler}
\author{D. Leibfried}
\author{D. J. Wineland}
\author{D. P. Pappas}
\affiliation{NIST, 325 Broadway, Boulder, Colorado 80305}
%

%===============
\begin{abstract}
Electric-field noise from the surfaces of ion-trap electrodes couples to the ion's charge causing heating of the ion's motional modes.  This heating limits the fidelity of quantum gates implemented in quantum information processing experiments.  The exact mechanism that gives rise to electric-field noise from surfaces is not well-understood and remains an active area of research.  In this work, we detail experiments intended to measure ion motional heating rates with exchangeable surfaces positioned in close proximity to the ion, as a sensor to electric-field noise.  We have prepared samples with various surface conditions, characterized in situ with scanned probe microscopy and electron spectroscopy, ranging in degrees of cleanliness and structural order.    The heating-rate data, however, show no significant differences between the disparate surfaces that were probed.  These results suggest that the driving mechanism for electric-field noise from surfaces is due to more than just thermal excitations alone.
\end{abstract}
%===============
%\pacs{}
\date{\today}
\maketitle

\section{Introduction}

Trapped ions are used to study intriguing physics of quantum mechanics, which has led to useful applications, such as precision spectroscopy [1], ultra-low force sensing [2], high-precision atomic clocks [3, 4], and quantum-logic gate operations for quantum information processing (QIP) [5, 6].  A key aspect of these experiments is the use of the internal states of the ions as quantum bits, coupled to the motional modes of the ions confined in the trap, which are often laser cooled to near the ground state of motion [7].  The internal state lifetimes are long and can be well-protected from outside influences.  However, the motional states are readily perturbed by environmental forces, e.g., from electric-field noise at the location and motional-mode frequency of the ions, typically 1 - 10 MHz.   This noise couples to the charge of the ions causing motional heating, where the ions uncontrollably acquire additional quanta, or phonons of motion.   Since the internal states and the motional states are coupled together to perform quantum logic gate operations, the rate at which the ions heat up can limit the fidelity of the operations.  Therefore, motional heating rates have become an important metric in the performance of ion traps used in QIP.

Various sources of noise can cause motional heating, however, one source is intrinsic to the trap electrodes themselves.  In fact, electric-field noise from the surfaces of the trap electrodes has proven to be a difficult problem to mitigate.  For decades, experimental evidence, based on the scaling of the heating rates with ion-electrode distance [8-10], electrode temperature [10-13], and motional-mode frequency [8-11, 13-16] has pointed to the surface of the electrodes as the source of the noise [17].  Moreover, the spectral density of the noise is typically orders of magnitude greater than that estimated to arise from typical noise sources from the bulk of the electrodes, e.g. from Johnson noise.  One working hypothesis has been that small independently fluctuating patches on the electrode surfaces are the source of the electric-field noise at the location of the ion, typically trapped 30 $\mu$m to 300 $\mu$m from the nearest electrode.   For QIP experiments, smaller traps are desired for scalability and faster gate speeds.  Because the heating rates scale strongly with the inverse of the ion-electrode distance, motional heating from surfaces presents a major obstacle to continued progress.

From a surface science perspective, one would naturally assume that adsorbed contaminants on the electrode surfaces play a role in this problem.  In fact, this was supported when heating rates in microfabricated ion traps were reduced by orders of magnitude after treatments of noble-gas ion bombardment [18-20].  Here, the assumption was that the reduction in noise was related to the removal of surface contaminants, however, the understanding of the root cause of electric-field noise from surfaces remains incomplete.

Various models have been put forth to explain these observations, providing insight into the problem.  One proposal models thermally excited vibrational modes of stationary adsorbate dipoles [21].  Others have considered the fluctuations of patches of various work functions, or the surface diffusion of adatoms as the source [10].  Recently, a first-principles study of dipole variations in carbon adatoms due to surface diffusion on a Au(110) surface has estimated a range for electric-field noise from this mechanism [22].  The results from this model are consistent with experimental values.

Our research focuses on experimentally determining the fundamental mechanism of this noise source.  We use traditional surface science tools to characterize the surface condition of various samples, in situ, with a novel ion trap designed to measure heating rates as a function of the ion-sample distance.  Since trapped ions are sensitive to the various conditions of the trap electrode surfaces [22], our experiment is designed to likewise measure noise from various sample surfaces.  In this paper, we describe experimental results that suggest that the driving mechanism for electric-field noise from surfaces is more than that due to thermal excitations alone.

% =============================
\section{Experimental Details} \label{motivation}
% =============================
Our experimental setup is comprised of an ultra-high vacuum (UHV) cluster, which includes facilities for sample fast-entry, preparation, and modification, scanned probe microscopy, and electron spectroscopy.  The vacuum system also houses a stylus-type ion trap [23, 24], with room temperature electrodes, and is equipped with a sample manipulator to position exchangeable sample surfaces within $\sim$ 50 $\mu$m of a trapped ion.
 We infer the electric-field noise spectral density $S_E$ at the location and motional frequency of an ion in the trap by measuring its heating rate $\dot{\overline{n}}$ (the time rate of change in the average number of quanta \textit{n} for a given motional mode) and using the following relation [8]:

\begin{equation}
S_E(\omega) = {4m \hbar \omega \over q^2} \dot{\overline{n}},
\label{eq:spectral noise}
\end{equation}

where $\omega/2\pi$ is the ion's motional frequency, \textit{q} is its charge, \textit{m} is its mass, and $\hbar$ is Planck's constant divided by $2\pi$ (also see Figure 2 in Ref. [17] for a brief description of the basic elements of a heating-rate measurement).  In the experiments described here, we trapped single $^{25}$Mg$^+$ ions 63 $\mu$m above the nearest electrode in a stylus-type Paul trap (in the absence of a test surface), similar to the trap described in Ref. [24].  We measured heating rates of a 4.4-MHz motional mode that is parallel to the sample surface (inset of Figure 3).

Since the electric-field noise from the trap itself adds a background to the measurement of noise from the sample surface, it is imperative to reduce the noise from the trap as much as possible.  The trap chips used in this work were microfabricated with electroplated Au electrodes.  The fabrication process leaves the electrode surfaces covered with several monolayers of contamination, which were removed by a pre-assembly treatment by Ne$^+$ bombardment detailed in Ref. [20].  After the final assembly of the trap in air and vacuum conditioning by baking, the level of surface contamination was greatly reduced compared to the as-fabricated surface with no Ne$^+$ treatment, and was shown to result in a significantly reduced electric-field noise spectral density of 5.7 $\times$ 10$^{-13}$ V$^2$m$^{-2}$Hz$^{-1}$ [20].  Finally, the trap was treated in situ with additional doses of Ne$^+$ bombardment to further reduce the background electric-field noise spectral density to 1.5 $\times$ 10$^{-13}$ V$^2$m$^{-2}$Hz$^{-1}$.  For comparison, untreated traps with room temperature electrodes and similar ion-electrode distances typically exhibit electric-field noise spectral densities one to two orders of magnitude higher than the above when normalized for the different motional frequencies used.  Over a period of 6 months in UHV, the background level of electric-field noise spectral density from the ion trap increased by a factor of 2.4, providing further evidence for the role of adsorbates.  After an additional in situ treatment by Ne$^+$ bombardment, a lower background noise level was recovered.  Heating rates were measured with samples in close proximity to the ion for both the high and low background cases.

A second, duplicate stylus-trap chip accompanied the ion trap used in this work through each step of the process, i.e. the preassembly treatment by Ne$^+$ bombardment, exposure to air, vacuum processing, and the additional in situ treatments.  The operating trap was not accessible for surface analysis after the final assembly, however the duplicate stylus-trap chip was transportable and analyzed with the various surface analysis tools [20] (also cf. Figure 2 below).  This afforded a comparison between the operating trap and the duplicate under the same surface-treatment conditions.

The samples that were brought into close proximity to the trapped ion were of various surface conditions with an increasing level of surface contamination:

\# 1) a Ne$^+$ sputter-treated electroplated-Au film with no detectable contaminants as determined by Auger electron spectroscopy (AES), with ~ 0.05 monolayer (ML) sensitivity,

\# 2) a Ne$^+$ sputter-treated Au(110) crystal (unannealed) with a submonolayer coverage of carbonaceous contamination resulting from long, post-treatment exposures to the background gas in UHV ($\sim$ 4 months),

\# 3) an untreated electroplated-Au film with approximately 2 equivalent ML of carbonaceous contamination, and

\# 4) an as-fabricated electroplated-Au surface-electrode ion-trap chip [18, 25] with approximately 3 equivalent ML of carbonaceous contamination.

The Auger spectra for each of the samples used in this work are shown in Figure 1.  Samples \# 1 - 3 were 1-mm diameter cylindrical posts, the geometry with respect to the stylus ion trap is shown below (cf. Figure 3 inset).  The Au films in samples \# 1 and \# 3 were electrodeposited on bulk Cu substrates using the same Au plating process as the stylus trap itself.   Sample \# 4 was microfabricated in the same process as the ion trap used in Ref. [18] and mounted such that the electrode structures were positioned near the ion during testing.  Each of the samples was electrically grounded or dc-biased outside of the vacuum chamber.  We are particularly interested in thick ($\sim$ 10 $\mu$m) electroplated Au films (untreated and sputter-treated) because they are used in the current fabrication process for surface-electrode traps at NIST and elsewhere [25, 26].    The use of thick electroplated Au electrodes with narrow inter-electrode gaps in surface-electrode traps helps to shield the trapped ion from stray fields associated with charging of the insulating substrate (typically crystalline quartz).
%======= Figure 1 ========================
\begin{figure}[h]
\begin{center}
\includegraphics[width=0.40\textwidth]{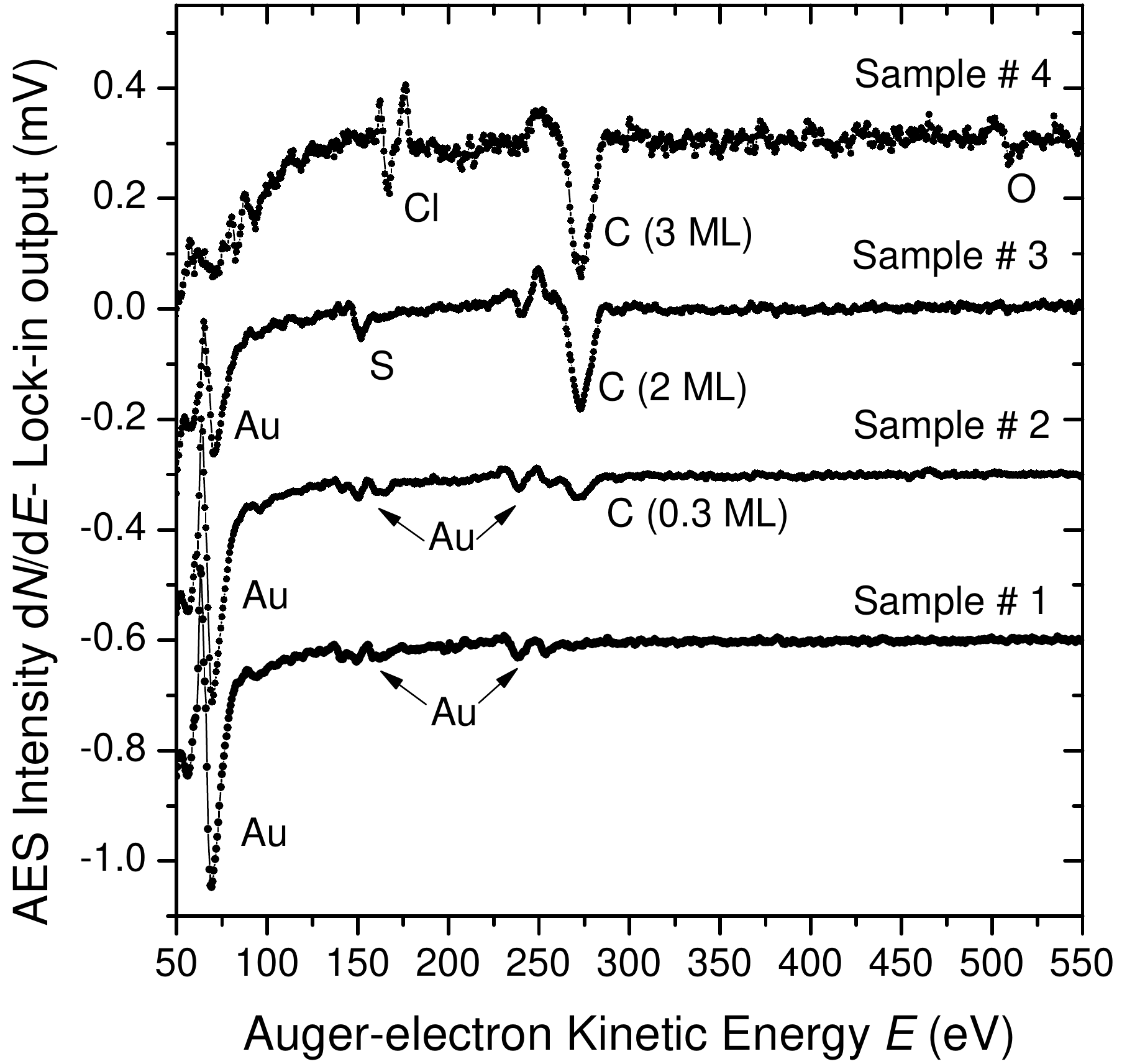}
\end{center}
\caption{\label{fig1-noisedata} Auger-electron spectra for samples \# 1 - 4.  The various degrees of carbon coverage were estimated by determining the ratio of the normalized intensities of the C and Au AES lines, accounting for attenuation due to the electron's inelastic mean free path.  Samples \# 3 and \# 4 also have small amounts of S and Cl, respectively, presumed to originate from the electroplating process.  The spectra are offset vertically for clarity.  The different surface conditions of the samples were intended to heat a trapped ion differently when positioned in close proximity.
}
\end{figure}
%======= Figure 1 ========================

% =============================
\section{Results and Discussion} \label{methods}
% =============================

The electric-field noise from the trap depends strongly on the condition of the electrode surfaces.  When treated either in situ or ex situ with ion bombardment, significantly lower electric-field noise spectral densities have been reported [18-20].  Moreover, it has been inferred that the strength of the noise behaves non-monotonically in the sub-monolayer coverage regime when incrementally removing the contaminants with ion bombardment [22].  Concomitant with the significant reduction in electric-field noise, in addition to the decreased concentration of surface impurities to undetectable levels using AES, we have also observed medium-range order over $\sim$ 100 nanometers after the ion-bombardment treatments.  Prior to treatment with ion bombardment, scanning tunneling microscopy (STM) images (not shown) reveal a clustered and disordered morphology, as one would expect from a contaminated surface.  After the ex situ pre-assembly treatment to the duplicate stylus-trap chip, ordered structures are observed as seen in STM, shown in Figure 2.  Figure 2(a) shows the typical range over which the order extends.  In Figure 2(b), atomic-scale order is seen to resemble the structure of Au(100), where the missing-row-like features are presumed to be due to a strain relieving mechanism.   Post-annealing of the sputter-treated surface was not implemented in our ion-trap chip mount, therefore over a long range, there was a rough hill-and-valley morphology due to the ion bombardment [27].  The possibility of surface order playing a role in electric-field noise from surfaces, in addition to their degree of cleanliness, is an intriguing avenue for future work.

%======= Figure 2 ========================
\begin{figure}[b]
\begin{center}
\includegraphics[width=0.40\textwidth]{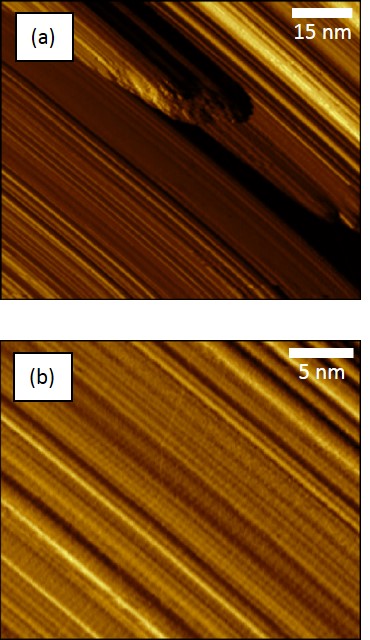}
\end{center}
\caption{\label{fig2-STM} Derivative STM images of the duplicate stylus trap after the pre-assembly treatment.  Figure 2(a), (90 $\times$ 74) nm$^2$, 15 nm full \textit{z} scale in topography, shows the typical extent of the ordering observed after ion bombardment.  Figure 2(b), (28 $\times$ 23) nm$^2$, 1.5 nm full \textit{z} scale in topography, details the atomic-scale order, resembling a Au(100)-like surface.   Because these surfaces were not annealed (only vacuum baked to 450 K), over long range the surfaces have a rough hill-and-valley morphology. }
\end{figure}
%======= Figure 2 ========================

The various conditions of our sample surfaces, described in the previous section, were intended to exhibit a wide range of electric-field noise spectral densities when positioned in close proximity to the trapped ion.  For example, sample \# 1, the sputter-treated clean Au film, was used to characterize the effect of placing a ground plane at various positions, only introducing minimal additional heating to the ion.   When the samples are placed close to the ion, they become an additional electrode in the trap as an rf ground.  As the sample is positioned closer to the ion, the rf null, where the ion is confined, is pushed closer to the trap as shown in Figure 3.  In other words, as the sample-to-stylus distance \textit{h} is reduced, both the ion-sample distance \textit{y} and the ion-stylus distance \textit{d} become smaller as well.  As a result, in this geometry with the distances used here, the ion is always closer to the trap electrode than the sample surface of interest.  Because the motional heating rate scales strongly with the inverse of the distance to the nearest electrode, the increased heating from decreasing \textit{d} is expected to be due to electric-field noise from the stylus trap itself, when using a treated sample with minimal additional heating.  Assuming this sample gives rise to negligible heating, the scaling of the heating rate as a function of the ion-stylus distance \textit{d} follows a power law, where $\dot{\overline{n}}$ $\sim$ $d^{-3.1}$ , as determined by a fit to the data.

%======= Figure 3 ========================
\begin{figure}[t]
\begin{center}
\includegraphics[width=0.40\textwidth]{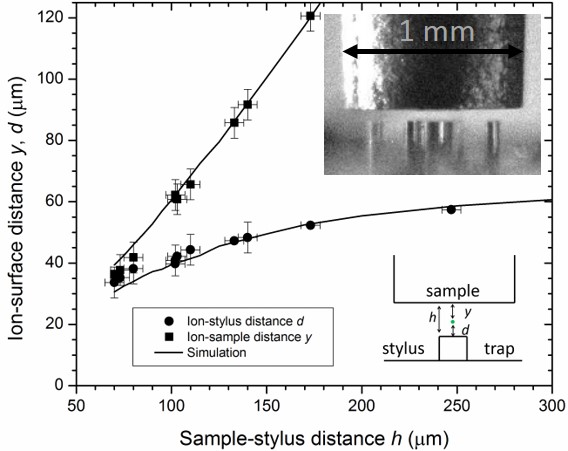}
\end{center}
\caption{\label{fig3}
Behavior of the ion-surface distances vs. the sample-stylus distance.  As the grounded samples are moved closer to the trap, the rf null, where the ion is confined, also moves closer to the trap.  The distance measurements agree well with results from a three-dimensional electrostatic simulation that included the ion-trap electrodes and sample post to determine the ion height.  The top inset is a side-view picture of the stylus-trap electrodes (200 $\mu$m tall) with a 1-mm diameter surface (sample \# 2) in close proximity.  Heating rates were measured on a ($^{25}$Mg$^+$) 4.4-MHz motional mode that is parallel to the sample surface.
}
\end{figure}
%======= Figure 3 ========================

Data from the heating rate measurements for the various samples used in this study are shown in Figure 4 as a function of the ion-surface distance.  For comparison, we show the ion-stylus distance \textit{d} on the bottom axis as well as the ion-sample distance \textit{y} on the top axis.  The open circles at $d = 63 \mu$m represent the heating rates for the bare stylus-trap background with no proximal sample, where the higher heating rate is for the trap condition after the 6-month period in UHV between sputter treatments.  The lower background heating rate (lower open circle at $d = 63 \mu$m) was measured after a final sputter treatment, and the lower heating-rate data are for sample \# 1, the cleaned sputter treated Au film, and sample \# 4, the as-fabricated ion-trap chip.  The ion-stylus distance used as the independent variable is the more relevant dimension, since the data for the various samples show no significant difference for the disparate surface conditions.  It would be expected, based on the assumption of a role played by thermally driven adsorbates, that samples \# 3 and \# 4 would causes a significant increase in the motional heating of the ion, especially when the samples are at comparable distances to the ion as to the trap electrode surfaces.  The fact that the samples with these different surface conditions do not show significantly different results suggests that the heating from these samples is negligible.  This finding also suggests that the driving mechanism for electric-field noise from surfaces is more than that due to thermal excitations alone.

%======= Figure 4 ========================
\begin{figure}[b]
\begin{center}
\includegraphics[width=0.40\textwidth]{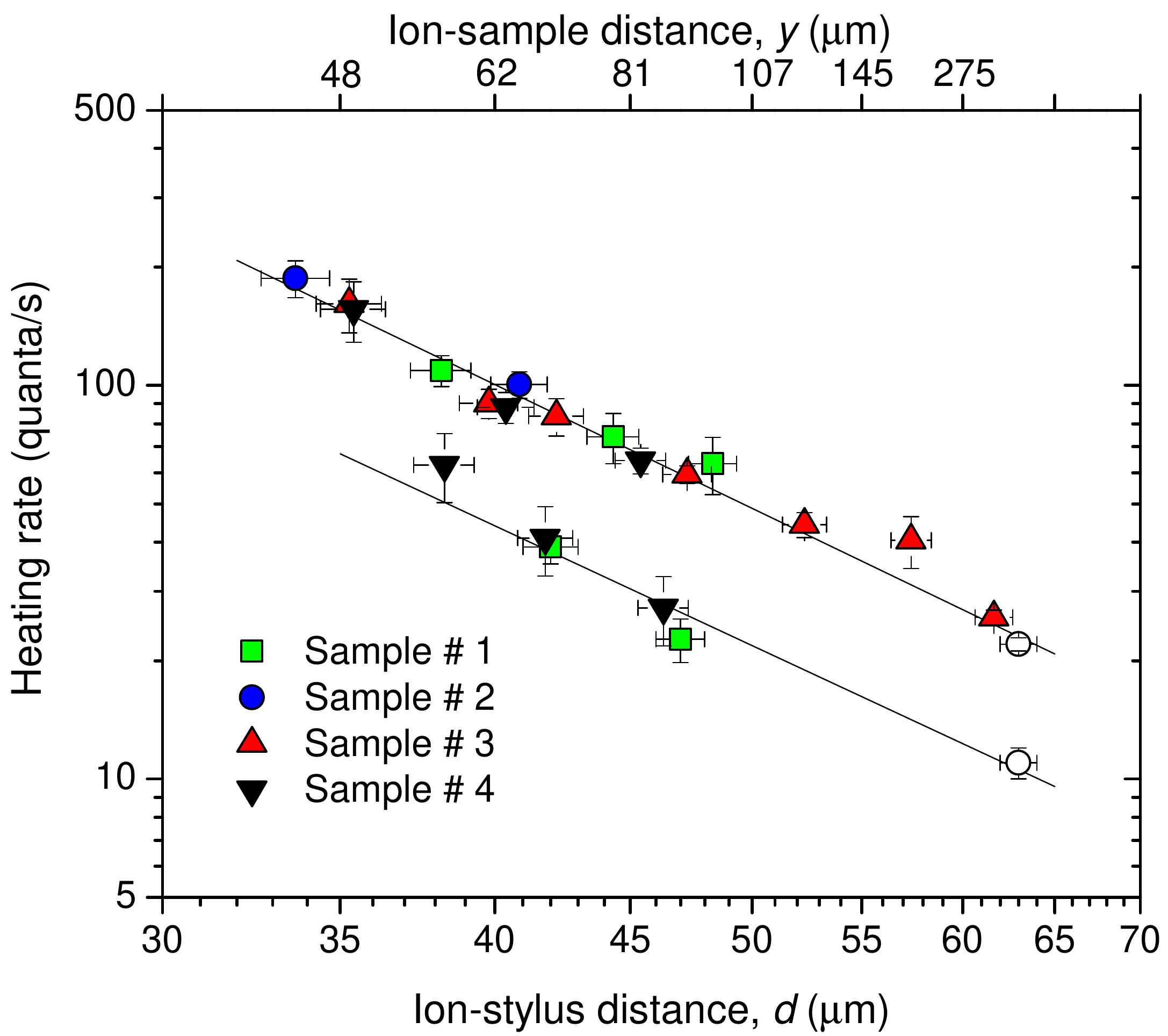}
\end{center}
\caption{\label{fig4} Heating rate data for samples \# 1 - 4 at various positions.  The upper set of data is for the higher background heating from the trap after 6 months in UHV following the initial sputter treatments.  The lower background data set was collected after a final sputter treatment to the trap.  The fact that the samples with different surface conditions do not show different results suggests that the heating from these samples is negligible.  The solid lines depict a $d^{-3.1}$ scaling, implying that the heating rates are dominated by electric-field noise from the trap.
}
\end{figure}
%======= Figure 4 ========================

The sensitivity of these measurements can be estimated making use of metadata from the literature [10].  When one considers heating-rate data from other untreated ion-trap experiments with ion-electrode distances comparable to our ion-sample distance, assuming similar surface conditions as our untreated samples, it is estimated that heating from our untreated samples should be 2 to 40 times above the background of the stylus trap over the range of our ion-sample distances [9, 12, 15, 28].  Of course, comparing to metadata in this way is not precise because there are assumptions made about how the heating rates scale with motional frequency and the unknown condition of the various electrode surfaces in the other traps.  Making such a comparison, however, does provide a level of confidence that our data should have shown a stronger dependence on ion-sample distance than that measured.

When considering the physical and electrical differences between typical untreated ion-trap electrodes, which are known to produce high levels of electric-field noise [18], and our sample surfaces, particularly the ion-trap chip, sample \# 4, which had the physical characteristics of an actual ion trap, the remaining distinction is the application of rf potentials to trap electrodes, which were absent from our samples.  Along these lines, one might consider how rf fields or currents at the surface drive adsorbate dipoles (or charge distributions) to fluctuate more than that due to thermal excitations alone, to enhance the effective temperature.  This idea, however, presents a conundrum because when the samples become a part of the rf trap when placed in close proximity, the rf fields should be present at their surfaces as well.  One explanation for this could be that the impedance due to the samples' electrical geometry limits the rf currents in our samples, and therefore affect the field distributions at the sample surfaces.  In any case, the results presented here challenge the conventional thinking that electric-field noise in ion traps is due only to thermally driven adsorbate processes on the electrode surfaces.

%===============================
\section{Conclusions}
%===============================
In summary, we have described the use of a novel surface-science/ion-trap apparatus to measure trapped-ion motional heating rates and infer the electric-field noise from proximal sample surfaces.  We have characterized the samples, which have various surface conditions, making use of in situ scanned probe microscopy and electron spectroscopy.  Because surface treatments by ion bombardment have previously been shown to reduce the electric-field noise in trapped-ion heating-rate measurements by orders of magnitude, some samples were sputter treated and others remained untreated.  The fact that there was no measurable difference in motional heating for the various samples suggests that electric-field noise in ion traps may be driven by more than thermal excitations.  Ongoing experiments are aimed at elucidating the role played by rf electric fields near surfaces, acting to induce an additional effective temperature, as a driving mechanism. Based on many other ion-trap heating-rate experiments, the sensitivity of our measurement is estimated to be sufficient.  These results show that motional heating in ion traps remains a complex problem in the field of surface science.
%================================================================

%============================
\section*{Acknowledgments}
%============================

We would like to thank C. Arrington and Sandia National Laboratories for providing the electroplated-Au samples and the fabrication of the stylus traps [24].  We also thank A. Wilson and D. Slichter for helpful discussions, and R. Lake and J. Bollinger for critical suggestions on the manuscript.  This article is a contribution of the U.S. Government and is not subject to U.S. copyright.

%========================

%====================

%=============
\end{document}